\documentclass[twocolumn,floats,showpacs,prl]{revtex4}
\usepackage[above]{placeins}
\usepackage{amsmath}
\usepackage{graphicx}

\begin{document}

\title{Optimum and quasi-optimum doped superconducting phases in FeAs and Fe$%
_{2}$As$_{2}$ superconductors}
\author{Xiuqing Huang$^{1,2}$}
\email{xqhuang@nju.edu.cn}
\affiliation{$^1$Department of Physics and National Laboratory of Solid State
Microstructure, Nanjing University, Nanjing 210093, China \\
$^{2}$ Department of Telecommunications Engineering ICE, PLAUST, Nanjing
210016, China}
\date{\today}

\begin{abstract}
We have developed a real space spin-parallel theory of superconductivity
based on the minimum principle in energy. This theory has successfully
provided coherent explanations to a number of complicated problems in
conventional and non-conventional superconductors. In this paper, we report
the study the optimal doping problem in the new oxypnictide high-temperature
superconductors using aforementioned theory. In FeAs family, it is shown
that there are three optimum (or quasi-optimum) doped phases at doping
levels $x_{1}=1/3$, $1/6$ and $1/8$, where the vortex lattice forms square
or triangular stable configurations. While in Fe$_{2}$As$_{2}$ family, the
optimal dopings occur at $x_{2}=2/5$ and $1/2$ with square and triangular
superconducting vortex line lattices, respectively.
\end{abstract}

 \maketitle

We should not be surprise to find that the iron-based compounds can become
superconductors \cite{kamihara,chen,rotter,chu}. In fact, superconductivity
is a very common physical phenomenon which may occur in any materials with
an appropriate charge carrier density (not too high, not too low). It is
also not surprising that the pseudogap state (localized Cooper pair) was
observed in the underdoped FeAs superconductors \cite{sato}. In my opinion,
pseudogap is also a common physical phenomenon in some systems where the
carrier concentrations are much more sparse and the pair-pair interactions
can be neglected \cite{huang0}.

In the face of various superconductors: cuprate, C$_{60}$, MgB$_{2}$, Sr$%
_{2} $RuO$_{4}$, FeAs, Fe$_{2}$As$_{2}$, etc., it is now clear that we need
to develop a unified theory of superconductivity \cite{huang1}. Without
Hamiltonian, without wave function, without electronic bands or orbitals,
without quantum field theory, the suggested theory has provided a beautiful
and consistent picture for describing the myriad baffling microphenomena
which had previously defied explanation. For example, the local checkerboard
patterns and \textquotedblleft magic doping fractions\textquotedblright\ in
La$_{2-x}$Sr$_{x}$CuO$_{4}$ \cite{huang1}, the tetragonal vortex phase in Bi$%
_{2}$Sr$_{2}$CaCu$_{2}$O$_{8}$ \cite{huang2}, the hexagonal vortex lattice
and charge carrier density in MgB$_{2}$ \cite{huang2}, the optimal doping
phases \cite{huang2}, pressure effects \cite{huang3} and pseudogap phase
\cite{huang0} in the new iron-based superconductors, and the $4a\times 4a$
and $4\sqrt{2}a\times 4\sqrt{2}a$ checkerboard patterns in hole-doped Ca$%
_{2-x}$Na$_{x}$CuO$_{2}$Cl$_{2}$ \cite{huang4}. The encouraging agreement of
our results with the experiments implies a possibility that our theory would
finally open a new window in condensed matter physics.

Although the proposed theory is consistent with many superconducting
experimental phenomenology, it is very difficult to persuade people that
such a simplistic picture would be the mechanism of Cooper pairing and
superconducting in the most complicated strong correlation systems.
Obviously, the condensed matter community is unaccustomed to a theory of
superconductivity without involving Hamiltonian and wave function. They
can't accept the fact that the Hamiltonian-based famous BCS theory is
physically incorrect and even doesn't work for the conventional
superconductors \cite{huang1,huang2}.

In the present paper, we try to extend the application of the theory to the
optimal doping problem in FeAs and Fe$_{2}$As$_{2}$ layered superconductors.
With this we aim to stress that our theory is based on the most solid
\textit{minimum energy principle}, not just arguments or groundless.
Physically, in a material, the dominant structural phase should be a
minimum-energy state which satisfies the basic symmetry of the crystal
structure. In this sense, the superconducting and non-superconducting states
are merely some minimum energy condensed states of the electronic charge
carriers, or some kinds of real-space low-energy charge orders (the
so-called Wigner crystals). In our viewpoint, the minimum-energy based
superconducting theory \cite{huang0,huang1,huang2,huang3,huang4} exhibits
remarkably the beauty and mystery of physics, moreover, this approach is the
most reliable one. Generally, a correct microscopic physical theory should
be established in the following way: (the fundamental law of nature: \textit{%
minimum energy principle)}$\rightarrow $(\textit{microscopic symmetry of
material structure})$\rightarrow $(\textit{physical or scientific laws). }%
Here we would like to reemphasize that a correct and reliable physical
theory must firstly be mathematically simple and non-approximate.\textit{\ }%
Time can prove that all the existing theories of superconductors based on
complex mathematical theorems are indeed on the wrong track.

\begin{figure}[tbp]
\begin{center}
\resizebox{1\columnwidth}{!}{
\includegraphics{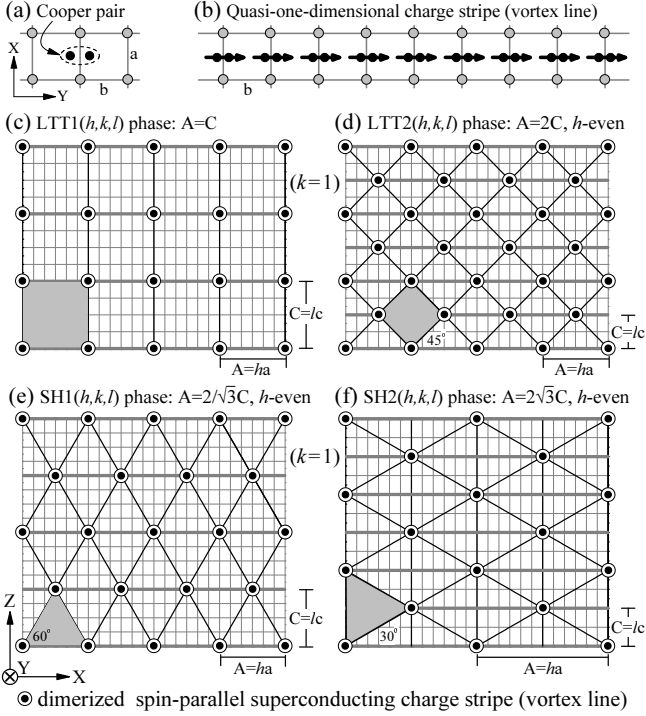}}
\end{center}
\caption{The schematic interpretation of the theory of superconductivity
based on the minimum energy principle in FeAs superconductors. (a) A
quasi-zero-dimensional localized Cooper pair, (b) a quasi-one-dimensional
dimerized vortex line, (c)-(f) four quasi-two-dimensional vortex lattices
with a uniform distribution of vortex lines.}
\label{fig1}
\end{figure}

Fig. \ref{fig1}(a) shows a real-space quasi-zero-dimensional localized
Cooper pair. In a previous paper \cite{huang0}, it has been shown that there
are two special positions where the localized Cooper pair will be in its
minimum energy states. We have argued that pairing in superconductors is an
individual behavior characterized by pseudogap, while superconductivity is a
collective behavior of many coherent electron pairs \cite{huang1}. To
maintain a stable superconducting phase (minimum energy), first the pairs of
electrons must condense themselves into a real-space quasi-one-dimensional
dimerized vortex line (a charge-Peierls dimerized transition), as shown in
Fig. \ref{fig1} (b). Second, the vortex lines can further self-organize into
some quasi-two-dimensional vortex lattices where a uniform distribution of
vortex lines is formed in the plane perpendicular to the stripes, as shown
in Figs. \ref{fig1} (c)-(f).

In the LTT1($h,k,l$) phase, as shown in Fig. \ref{fig1}(c), the charge
stripes have a tetragonal symmetry in XZ plane in which the superlattice
constants satisfy
\begin{equation}
\frac{A}{C}=\frac{ha}{lc}=1.  \label{lattice0}
\end{equation}%
Fig. \ref{fig1}(d) shows the LTT2($h,k,l$), the vortex lattice has a
tetragonal symmetry in XZ plane with a orientation 45$^{\text{0}}$ and the
superlattice constants:
\begin{equation}
\frac{A}{C}=\frac{ha}{lc}=2.
\end{equation}

While in simple hexagonal (SH) phases, as shown in Figs. \ref{fig1}(e) and
(f), the charge stripes possess identical trigonal crystal structures. In
the SH1($h,k,l$) phase [see Fig. \ref{fig1}(e)], the superlattice constants
have the following relation
\begin{equation}
\frac{A}{C}=\frac{ha}{lc}=\frac{2\sqrt{3}}{3}\approx 1.15470.
\end{equation}%
For the SH2($h,k,l$) phase of Fig. \ref{fig1}(f), this relation is given by
\begin{equation}
\frac{A}{C}=\frac{ha}{lc}=2\sqrt{3}\approx 3.46410.  \label{lattice}
\end{equation}

The appearance of the stable vortex lattices of Fig. \ref{fig1}\ is a common
feature of the optimally doped superconducting phases. This in turn leads to
some minimum-energy superconducting vortex phases with the highest
superconducting transition temperatures of the corresponding superconductors.

\begin{figure}[tbp]
\begin{center}
\resizebox{1\columnwidth}{!}{
\includegraphics{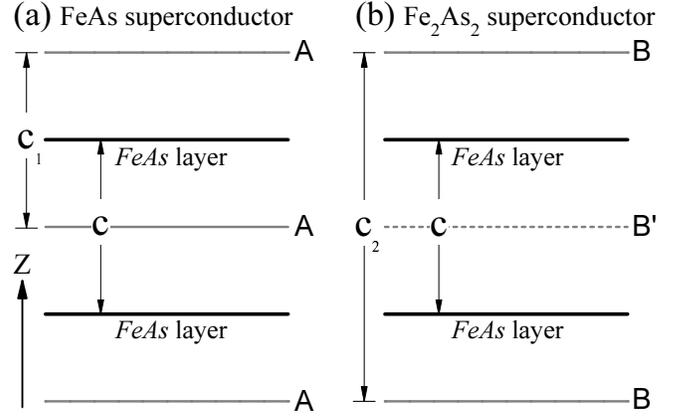}}
\end{center}
\caption{The schematic plot of the iron-based superconductors. (a) The FeAs
compounds, (b) the Fe$_{2}$As$_{2}$ compounds. }
\label{fig2}
\end{figure}

\begin{table*}[tbp]
\caption{Lattice constants, optimal and quasi-optimal superconducting
charge-stripe phases (vortex lattices) and doping levels (analytical values $%
x=2/hkl)$, $x_{1}=x$ and $x_{2}=2x$ for FeAs and Fe$_{2}$As$_{2}$ based
superconductors, respectively. Here $a=a_{0}\protect\sqrt{2}/2$ is the Fe-Fe
distance in two-dimensional square Fe lattices, where $a_{0}$ is the lattice
constant. The iron layer's distances are $c=c_{1}$ and $c_{2}/2$ for FeAs
and Fe$_{2}$As$_{2}$ compounds, respectively. And $\protect\xi _{xz}$ is the
minimum stripe-stripe separation.}
\label{table1}%
\begin{tabular}{cccccccccccccc}
\hline\hline
Superconductors & $a($\AA $)$ & $c($\AA $)$ & $h$ & $k$ & $l$ & $A/C$ & $%
(A/C)_{0}$ & $x$ & $x_{1}=x$ & $x_{2}=2x$ & Vortex phase & $\delta (\%)$ & $%
\xi _{xz}($\AA $)$ \\ \hline\hline
LaO$_{1-x}$F$_{x}$FeAs & $2.850$ & $8.739$ & 6 & 1 & 1 & 1.957 & 2 & $%
1/3\approx 0.333$ & $1/3$ &  & LTT2 & 2.15 & 12.360 \\
&  &  & 6 & 1 & 2 & 0.978 & 1 & $1/6\approx 0.167$ & $1/6$ &  & LTT1 & 2.15
& 17.478 \\
&  &  & 7 & 1 & 2 & 1.141 & $2/\sqrt{3}$ & $1/7\approx 0.143$ & $1/7$ &  &
forbidden & 1.19 &  \\
&  &  & 8 & 1 & 2 & 1.304 & $2/\sqrt{3}$ & $1/8=0.125$ & $1/8$ &  & SH1 &
12.7 & 22.800 \\ \hline
SmO$_{1-x}$F$_{x}$FeAs & $2.788$ & $8.514$ & 6 & 1 & 1 & 1.965 & 2 & 1/3 & $%
1/3$ &  & LTT2 & 1.75 & 12.041 \\
&  &  & 6 & 1 & 2 & 0.983 & 1 & 1/6 & $1/6$ &  & LTT1 & 1.75 & 16.728 \\
&  &  & 7 & 1 & 2 & 1.146 & $2/\sqrt{3}$ & 1/7 & $1/7$ &  & forbidden & 0.75
&  \\
&  &  & 8 & 1 & 2 & 1.309 & $2/\sqrt{3}$ & 1/8 & $1/8$ &  & SH1 & 13.3 &
22.304 \\ \hline
Cs$_{1-x}$Sr$_{x}$Fe$_{2}$As$_{2}$ & $2.765$ & $13.760/2=6.880$ & 5 & 1 & 2
& 1.004 & 1 & $1/5=0.20$ &  & $0.4$ & LTT1 & 0.40 & 13.825 \\
&  &  & 8 & 1 & 1 & 3.215 & $2\sqrt{3}$ & $1/4=0.25$ &  & $0.5$ & SH2 & 7.19
& 13.760 \\ \hline
K$_{1-x}$Sr$_{x}$Fe$_{2}$As$_{2}$ & $2.751$ & $12.948/2=6.474$ & 5 & 1 & 2 &
1.062 & 1 & 1/5 &  & $0.4$ & LTT1 & 6.20 & 13.755 \\
&  &  & 8 & 1 & 1 & 3.399 & $2\sqrt{3}$ & 1/4 &  & $0.5$ & SH2 & 1.78 &
12.948 \\ \hline
Ba$_{1-x}$K$_{x}$Fe$_{2}$As$_{2}$ & $2.764$ & $13.212/2=6.606$ & 5 & 1 & 2 &
1.046 & 1 & 1/5 &  & $0.4$ & LTT1 & 4.60 & 13.820 \\
&  &  & 8 & 1 & 1 & 3.347 & $2\sqrt{3}$ & 1/4 &  & $0.5$ & SH2 & 3.37 &
13.212 \\ \hline\hline
\end{tabular}%
\end{table*}

After the initial report of $T_{c}=26$ K in LaO$_{1-x}$F$_{x}$FeAs \cite%
{kamihara}, a new class of FeAs-materials with a promising potential for
higher $T_{c}$ has been discovered \cite{chen}. The structure of these
layered compounds is sketched in Fig. \ref{fig2} (a), where the FeAs layers
may contribute to the superconductivity and the A layers are the
charge-reservoirs. Figure \ref{fig2} (b) presents a crystal structure of the
new Fe$_{2}$As$_{2}$ family where there are two FeAs superconducting layers
within a unit cell of the superconductors, the superconducting currents flow
in the FeAs layers, here, the B and B' layers are non-superconducting
charge-reservoirs. Some researchers considered that the new iron-based
superconductors disclose a new physics, contain new mysteries and may start
us along an uncharted pathway to room temperature superconductivity. But we
think it is not the appropriate time for us to talk about the room
temperature superconductors. What we are more concerned about at present is:
which physical parameters play an important role in raising $T_{c}$ of the
superconductors? According to our theory of Fig. \ref{fig1}, $T_{c}$ can be
tuned directly by varying the lattice constants and charge carrier density
of the superconductors. In other words, for a given superconductor, there
exist some optimal matching conditions between the lattice constants and
charge carrier density which support the highest superconducting transition
temperature of the superconductor, or the optimal doping problem.

Theoretical and numerical studies have shown that superconductivity in FeAs
and Fe$_{2}$As$_{2}$ compounds is associated with the two-dimensional square
Fe layers. In the framework of the minimum energy of the pairing and
superconducting mechanism (see Fig. \ref{fig1}), two types of iron-based
superconductors (see Fig. \ref{fig2}) are essentially the same. Based on the
experimental data of lattice constants, we obtain the relation between
lattice constants and optimal (or quasi-optimal) vortex phases in the FeAs
and Fe$_{2}$As$_{2}$ superconductors, as shown in Table \ref{table1}. A new
structural parameter $\delta $ used for evaluating the vortex lattice
deformation is given by
\begin{equation*}
\delta =\frac{\left\vert A/C-(A/C)_{0}\right\vert \times 100}{(A/C)_{0}}\%,
\end{equation*}%
where $(A/C)_{0}$ is one of the analytical values of Eqs. (\ref{lattice0})-(%
\ref{lattice}) and the $A/C$ is the corresponding numerical result estimated
on the basis of the experimental values of the lattice constants. Normally,
the higher the $\delta $ is, the more serious the vortex lattice
deformation, as a consequence, the corresponding superconducting phase may
be less stable and exhibit a lower superconducting transition temperature.
In addition, a large $\delta $ value at the same time means a stronger
pressure effect on superconductivity in the superconductor.

\begin{figure}[tbp]
\begin{center}
\resizebox{0.57\columnwidth}{!}{
\includegraphics{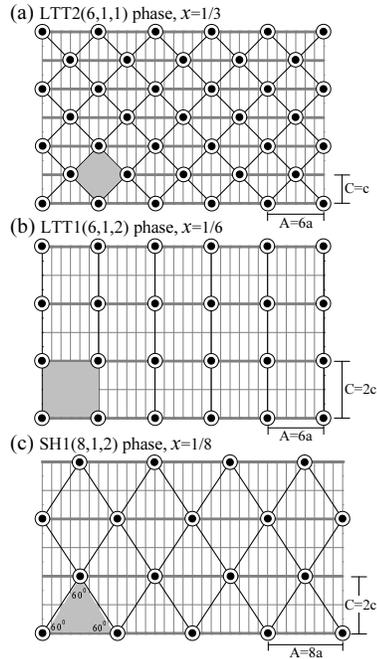}}
\end{center}
\caption{{}A schematic description of optimal and quasi-optimal
superconducting vortex phases in FeAs superconductor. (a) A crowded
quasi-optimal vortex phase at $x_{1}=1/3$ with a small stripe-stripe
separation $\protect\xi _{xz}$, (b) the optimal doped phase at $%
x_{1}=1/6=0.167$, and (c) a serious distorted quasi-optimal vortex phase at $%
x_{1}=0.125$ with a large $\protect\delta $ value.}
\label{fig3}
\end{figure}

For the FeAs superconductors, as shown in Table \ref{table1} of LaO$_{1-x}$F$%
_{x}$FeAs and SmO$_{1-x}$F$_{x}$FeAs, the analytical results indicate that
the candidates for the optimal doping vortex phases are LTT2($6,1,1$), LTT1($%
6,1,2$) and SH1($8,1,2$), respectively. And the corresponding optimal doping
levels are $x_{1}=1/3\approx 0.333$ \cite{gao}, $1/6\approx 0.167$ \cite%
{chen} and $1/8=0.125$ \cite{kamihara}, respectively. These three vortex
phases are shown in Fig. \ref{fig3}. From these results, it becomes evident
that the LTT1($6,1,2$) of Fig. \ref{fig3} (b) with a small quantity of $%
\delta \sim 2\%$ and an appropriate stripe-stripe separation $\xi _{xz}\sim
17\mathring{A}$ is the optimal doped superconducting vortex phase, while the
LTT2($6,1,1$) of Fig. \ref{fig3} (a) (a crowded vortex lattice with a small $%
\xi _{xz}\sim 12\mathring{A}$) and SH1($8,1,2$) of Fig. \ref{fig3} (c) (a
serious distorted vortex lattice with a large $\delta \sim 13\%$) are the
quasi-optimal superconducting phases. The hydrostatic-pressure effects on
the superconducting transition temperature of the LaO$_{1-x}$F$_{x}$FeAs ($%
x=0.11$) and SmO$_{1-x}$F$_{x}$FeAs ($x=0.13$) sample have been recently
reported by three research groups. \cite{wlu,takahashi,lorenz} These results
corroborate the suggested external pressure-induced $T_{c}$-enhancement in
the compound \cite{huang3}. It should be pointed out that the three samples
lie in the underdoped region with some distorted vortex lattices (for
example, SH1($9,1,2$) at $x=1/9=0.11$ and SH1($8,1,2$) at $x=1/8=0.125$), in
favor of the positive pressure effect on $T_{c}$.

\begin{figure}[tbp]
\begin{center}
\resizebox{0.85\columnwidth}{!}{
\includegraphics{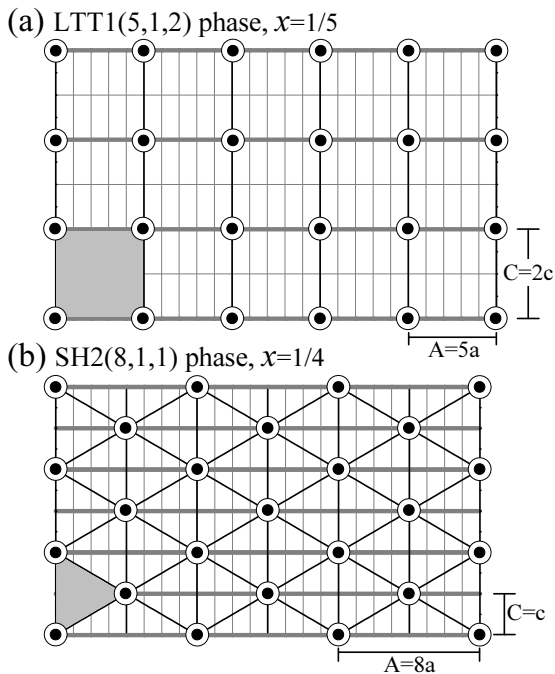}}
\end{center}
\caption{{}Two possible optimal doped superconducting vortex lattices in Fe$%
_{2}$As$_{2}$ superconductors. (a) The square LTT1($5,1,2$) at $%
x_{2}=2/5=0.4 $, and (b) the triangular SH2($8,1,1$) at $x_{2}=1/2=0.5.$}
\label{fig4}
\end{figure}

For the Fe$_{2}$As$_{2}$ superconductors \cite{rotter,chu}, there are two
competing optimal doped superconducting phases, they are LTT1($5,1,2$) of
Fig. \ref{fig4} (a) and SH2($8,1,1$) of Fig. \ref{fig4} (b) with the doping
levels $x_{2}=2/5=0.4$ and $1/2=0.5$, respectively. Experimental results for
Fe$_{2}$As$_{2}$ layered compounds show that the optimum doping occurs at $%
x_{2}$ approximately $0.4$ \cite{rotter} or $0.5$ \cite{chu},\ which are in
good agreement with our analytical results above.

In summary, the optimal doping problem in the new iron-based
high-temperature superconductors has been studied by using the newly
developed unified theory of superconductivity. In FeAs family, it is shown
that the optimum doped phase is LTT1($6,1,2$) at doping levels $x_{1}=1/6$,
where the vortex lattice forms square stable superconducting configuration.
Two quasi-optimal doped phases are also analytically determined, they are
the square vortex phase of LTT2($6,1,1$) at $x_{1}=1/3$ and the triangular
vortex phase of SH1($8,1,2$) at $x_{1}=1/8$. While in Fe$_{2}$As$_{2}$
family, the theoretical results show that two candidate optimal doping
phases are LTT1($5,1,2$) at $x_{2}=2/5$ and SH2($8,1,1$) at $x_{2}=1/2$ with
square and triangular superconducting vortex line lattices, respectively.

\end{document}